\documentclass[amstex,12pt,thmsa,sw20lart]{article}
\usepackage{amsfonts}
\usepackage{amsmath}
\usepackage{amssymb}

\input{tcilatex}
\begin{document}

\author{Piotr WILCZEK}
\title{Model-Theoretic Investigations into Consequence Operation (Cn) in
Quantum Logics: An Algebraic Approach. }
\date{}
\maketitle

\begin{abstract}
In this paper, we present the fundamentals of the so-called algebraic
approach to propositional quantum logics. We define the set of formulae
describing quantum reality as a free algebra freely generated by the set of
quantum proportional variables. We define the general notion of logic as a
structural consequence operation. Next, we introduce the concept of logical
matrices understood as a model of quantum logics.We give the definitions of
two quantum consequence operations defined in these models.
\end{abstract}

{\small \underline{2000 Mathematics Subjects Classification:} 03G12, 81P10.}


\bigskip

$\mathbf{1}$. \textbf{INTRODUCTION.}

\bigskip

Historically speaking we can distinguish two different and competitive ways
of understanding of the concept of \textquotedblleft \textit{logic}%
\textquotedblright . An approach considering the logic as a set of logically
valid sentences was the first manner of understanding logic. In this
approach one can perceive a logical system as a set of sentences closed
under substitutions and some rules of inference. A paradigmatic example is a
set of tautologies of classical propositional calculus. Second and more
general approach enables one to comprehend a logic as a \textit{logical
consequence operation} (or relation). This approach formalizes the most
general principles of reasoning and not a set of logically valid sentences.
Following the second approach one will uniquely obtain a set of logically
valid sentences as a set of consequences of an empty set of premises.
Following the first approach, i.e., starting from a set of logically valid
sentences one will not obtain a uniquely determined consequence operation.
So, there usually exist plenty of consequence operations for a given logical
system.

Summing up above considerations one can claim that logical validity does not
determine the rules of reasoning. Hence, the notion of logic can be
understood as a structural consequence operation discussed in detail in
section $3$.

In the literature concerning quantum logic there are only several articles
dealing with quantum logic as a structural consequence operation. In the
opinion of many logicians the notion of logic as a structural consequence
operation is one of the most important logical concepts. Considering logic
as a structural consequence operation belongs to the heritage of the \textit{%
Lvov-Warsaw School of Logic} and constitutes the basis for the development
of so-called \textit{Abstract Algebraic Logic} \cite{Fo}. The process of an 
\textit{algebraization} of the logical system is rooted in the belief that
this process allows us to investigate general properties of logical systems
by stipulating that these properties are reflected in the properties of the
corresponding classes of algebras.

The idea of a logical calculus based on the relation between the properties
of a physical system and the self-adjoint projection operators defined on a
Hilbert space can be traced back to the work of J. von Neumann \cite{Bir} .

In our papers we follow the so-called \textit{Geneva-Brussels Approach} to
the foundations of quantum physics \cite{Ae,Sme}. This approach can be
alternatively termed \textit{Operational Quantum Logic} \cite{Sme} and
corresponds to the theory of \textit{Property Lattices}.\ The general idea
of operational quantum logic is to give a complete formal description of
physical systems in terms of their actual and potential properties and a
dual description in terms of their states. Fundamental notion of quantum
logic is that of \textquotedblleft \textit{yes-no}\textquotedblright\
question\ or \textquotedblleft \textit{definite experimental project}%
\textquotedblright . A \textquotedblleft yes-no\textquotedblright\ question $%
\alpha \in Q$ is an experimental procedure and can be understood as a list
of concrete actions accompanied by a rule that specifies in advance with
outcomes count a positive response. A question is named \textquotedblleft 
\textit{true}\textquotedblright\ for a particular physical system if it is
certain that \textquotedblleft \textit{yes}\textquotedblright\ would be
obtained when the experimental procedure is performed, and is called
\textquotedblleft \textit{false}\textquotedblright\ otherwise \cite{Sme}.
The main point being that the structure of mathematical representatives for
experimental propositions of a quantum system, corresponding to the
projections on a Hilbert space forms an orthomodular lattice - or
equivalently -- can be modeled by orthomodular lattices. Quantum logics
(just like classical logic) are a kind of propositional logic. They are
determined by a class of algebras. These algebras are defined by a set of
identities. In other words, each logic is formalized by a set of axiom
schemes and inference rules which correspond to its defining set of
identities. These logics represent a natural logical abstraction from the
class of all Hilbert space lattices. They are represented respectively by
orthomodular quantum logic $(OML)$ and by the weaker orthologic $(OL)$ which
for a long time has been also termed \textit{minimal quantum logic}.

This article tries to define two different notions of quantum consequence
operations: the \textit{weak} one and the \textit{strong} one $($section $3)$%
. In order to do that we must define the quantum sentential calculus as an
absolutely free algebra $($section $2)$. We will give full model-theoretic
characterization of quantum logic which enables us to define two quantum
consequence operations $($section $4)$.\medskip \bigskip 

$\mathbf{2}$. \textbf{PRELIMINARY REMARKS.}

\bigskip

Every algebra we consider here has the signature $\left\langle \mathbf{A}%
,\leq ,\cap ,\cup ,(\cdot )^{\prime },\mathbf{0},\mathbf{1}\right\rangle $
and is of similarity type $\left\langle 2,2,1,0,0\right\rangle $. Algebraic
structures, in particular algebras, will be labeled with set of boldface
complexes of letters beginning with a capitalized Latin characters, e.g., $%
\mathbf{A}$, $\mathbf{B}$, $\mathbf{Fm}$, \ldots , and their universes by
the corresponding light-face characters, $A$, $B$, $Fm$, \ldots . All our
classes of algebra are \textit{varieties} $($we define variety as a
equationally definable class of algebra$)$. The varieties of ortholattices
is denoted by $\mathbf{OL}$. In order to show that this class constitutes a
variety explicitly, we give its definition by the set of identities:\bigskip

\textbf{Definition }$\mathbf{1}$. An \textit{ortholattice} is an algebraic
structure $\mathcal{A=}\left\langle \mathbf{A},\leq ,\cap ,\cup ,(\cdot
)^{\prime },\mathbf{0},\mathbf{1}\right\rangle $ which satisfies the
following identities:%
\begin{eqnarray*}
x\cap y &=&y\cap x. \\
x\cap (y\cap z) &=&(x\cap y)\cap z. \\
x &=&x\cap (x\cup y).
\end{eqnarray*}%
\begin{eqnarray*}
x\cup y &=&y\cup x. \\
x\cup (y\cup z) &=&(x\cup y)\cup z. \\
x &=&x\cup (x\cap y).
\end{eqnarray*}%
\begin{eqnarray*}
x\cup \mathbf{1} &=&\mathbf{1}. \\
x\cap x^{\prime } &=&\mathbf{0}. \\
(x^{\prime })^{\prime }\cap x &=&x. \\
x^{\prime }\cap (x\cup y)^{\prime } &=&(x\cup y)^{\prime }.
\end{eqnarray*}

In other words, an ortholattice is a bounded lattice with a unary operation $%
(\cdot )^{\prime }$ which satisfies the following: for any $x,y\in A$

$a)$ $x\leq x^{\prime \prime }.$

$b)$ $x\cap x^{\prime }=\mathbf{0}.$

$c)$ $x\leq y$ implies $y^{\prime }\leq x^{\prime }.\bigskip $

The variety of $\mathbf{OML}$ of all \textit{orthomodular lattices}, the
class $\mathbf{MOL}$ of all \textit{modular ortholattices} and the class $%
\mathbf{BA}$ of all \textit{Boolean algebras} are defined by adding the 
\textit{orthomodular law}, the \textit{modular law} and the \textit{%
distributive law} respectively, to the identities for $\mathbf{OL}$.

One can represent it as follows:%
\begin{eqnarray*}
\text{For }\mathbf{OML}\text{ }x\cap \left\{ (x\cap y)\cup x^{\prime
}\right\} &=&x\cap y\text{ (orthomodular law).} \\
\text{For }\mathbf{MOL}\text{ }x\cap \left\{ (x\cap y)\cup z\right\}
&=&(x\cap y)\cup (x\cap z)\text{ (modular law).} \\
\text{For }\mathbf{BA}\text{ }x\cap (y\cup z) &=&(x\cap y)\cup (x\cap z)%
\text{ (distributive law).}
\end{eqnarray*}

\bigskip All classes, we mention here are varieties being subvarieties of $%
\mathbf{OL}$, and the relation between these varieties is:%
\begin{equation*}
\mathbf{BA}\subseteq \mathbf{MOL}\subseteq \mathbf{OML}\subseteq \mathbf{OL}.
\end{equation*}

Undoubtedly, there are many other subvarieties of $\mathbf{OL}$ not
mentioned here. In this introductory exposition we adopt a framework of
binary logic introduced by Goldblatt \cite{Gol}. First, we define the system
for a binary logic, which corresponds to the $\mathbf{OL}$ variety, and then
we extend this system by introducing several axiom schemes.

\bigskip

\textbf{Definition} $\mathbf{2}$. An \textit{orthologic} $\mathbf{OL}$ on
the set of formulae includes the following axioms and is closed under the
following inference rules:%
\begin{eqnarray*}
\text{Axiom schemes} &\text{:}& \\
(\text{Ax }1)\text{ }\alpha &\vdash &\alpha . \\
(\text{Ax }2)\text{ }\alpha &\vdash &\lnot \lnot \alpha . \\
(\text{Ax }3)\text{ }\alpha \wedge \beta &\vdash &\alpha . \\
(\text{Ax }4)\text{ }\alpha \wedge \beta &\vdash &\beta . \\
(\text{Ax }5)\text{ }\alpha &\vdash &\alpha \vee \beta . \\
(\text{Ax }6)\text{ }\beta &\vdash &\alpha \vee \beta . \\
(\text{Ax }7)\text{ }\alpha \wedge \lnot \alpha &\vdash &\beta . \\
(\text{Ax }8)\text{ }\lnot \lnot \alpha &\vdash &\alpha .
\end{eqnarray*}

\begin{eqnarray*}
\text{Inference rules} &\text{:}& \\
&&(\text{R }1)\text{ }\frac{\alpha \vdash \beta \text{ \ }\beta \vdash
\alpha }{\alpha \vdash \gamma }. \\
&&(\text{R }2)\text{ }\frac{\alpha \vdash \beta \text{ \ }\alpha \vdash
\gamma }{\alpha \vdash \beta \wedge \gamma }. \\
&&(\text{R }3)\text{ }\frac{\alpha \vdash \gamma \text{ \ }\beta \vdash
\gamma }{\alpha \vee \beta \vdash \gamma }. \\
&&(\text{R }4)\text{ }\frac{\alpha \vdash \beta }{\lnot \beta \vdash \lnot
\alpha }.
\end{eqnarray*}%
\bigskip

Subsequent logics are defined by adding additional axiom schemes:%
\begin{eqnarray*}
\text{the orthomodular logic }(OML)\text{ }\alpha \wedge (\lnot \alpha \vee
(\alpha \wedge \beta )) &\vdash &\beta . \\
\text{the modular orthologic }(MOL)\text{ }\alpha \wedge ((\alpha \wedge
\beta )\vee \gamma ) &\vdash &(\alpha \wedge \beta )\vee (\alpha \wedge
\gamma ). \\
\text{the classical logic }(CL)\text{ }\alpha \wedge (\beta \vee \gamma )
&\vdash &(\alpha \wedge \beta )\vee (\alpha \wedge \gamma ).
\end{eqnarray*}

The relation between strengths of these logics is shown below:%
\begin{equation*}
OL\rightarrow OML\rightarrow MOL\rightarrow CL\rightarrow \text{inconsistent
logics.}
\end{equation*}

In considering propositional quantum logic, we follow the path taken by
algebraically oriented logicians. We define a sentential language as an
absolutely free algebra. As a consequence of such definition we can
adequately describe basic properties of the propositional language \cite{Fo}.

First, we introduce the notion of the \textit{algebra of formulae} and we
denote it by $\mathbf{Fm}$. $\mathbf{Fm}$ is \textit{absolutely free algebra}
of type $\mathcal{L}$ over a denumerable set of generators $%
Var=\{p,q,...,r\} $. The set of generators - $Var$ - is identified with the
countable infinite set of propositional variables. The universe of $\mathbf{%
Fm}$ algebra is formed of inductively defined formulae. The set of formulae
describing quantum entity is inductively defined as the least set satisfying
the following conditions:\bigskip

$1)$ $Var\subset Fm$ where $Var=\left\{ p,q,...,r\right\} $ is the set of
quantum propositional variables.

$2)$ if $p,q,...,r\in Fm$ then finite sequence $F_{i}pqr$ also belongs to $%
Fm $ for any $i=1,2,...,n.\bigskip $

The $\mathbf{Fm}$ algebra is endowed with finitely many finitary operations $%
F_{1},F_{2},...,F_{n}$. Thus, $\mathbf{Fm}$ consists in the set of formulae
together with the operations of forming complex formulae associated with
each connective. The structure $\mathbf{Fm}=\left\langle
Fm,F_{1},F_{2},...,F_{n}\right\rangle $ is called the algebra of formulae -
or equivalently - the algebra of terms. The similarity type $\mathcal{L}$ of
the algebra depends on the number and arity of connectives.

The definition of language as a free algebra allows us to treat sentential
connectives as algebraic operations. The process of formation of complex
propositions from atomic ones is the algebraic process occurring between
elements of a given algebra.

\bigskip

$\mathbf{3}$. \textbf{CONSEQUENCE OPERATION AND LOGICS.}

\bigskip

In $1930$, Tarski defined what later on was called finitary consequence
operation - $Cn$ \cite{Tar}. A consequence operation is a particular case of
a closure operation \cite{Bur}. Consequence operation is a structural
consequence operation defined on the algebra of formulae if $Cn$ satisfies
the following conditions \cite{Tar,Fo}:\bigskip

$1)$ $X\subseteq Cn(X)$ reflexivity,

$2)$ if $X\subseteq Y$ then $Cn(X)\subseteq Cn(Y)$ monotonicity,

$3)$ $Cn(Cn(X))\subseteq Cn(X)$ idempotency,

$4)$ $eCn(X)\subseteq Cn(e(X))$ structurality.\bigskip

The last condition says that $Cn$ is closed with respect to \textit{%
substitutions} i.e., $Cn$ is \textit{substitution-invariant}. Algebraically
speaking, substitutions occurring in the algebra of terms can be understood
as an \textit{endomorphism} of these formulae. Substitutions in the
sentential language are defined as functions from a set of sentential
variables into the set of formulae. Formally, a substitution is the function%
\begin{equation*}
e:Var\rightarrow Fm.
\end{equation*}

Based on the fact, that the algebra of terms is the free algebra the
function $e$ can be extended to an endomorphism:%
\begin{equation*}
h^{e}Fm\rightarrow Fm.
\end{equation*}

Additionally, if $Cn$ satisfies the following condition:\bigskip

$5)$ $Cn(X)=\dbigcup \left\{ Cn(Y):Y\subseteq X,Y\text{ is finite}\right\}
\bigskip $

it is called a \textit{finitary} consequence operation.

A consequence operation $Cn$ on a set of formulae can be easily transformed
into a consequence relation $\vdash _{Cn}\subseteq $ $\mathcal{P}(Fm)\times
Fm$ between subsets of $Fm$ and elements of $Fm$ by postulating for every $%
X\subseteq Fm$ and every $\alpha \in Fm$ that%
\begin{equation*}
X\vdash _{Cn}\alpha \text{ if and only if }\alpha \in Cn(X)
\end{equation*}

where $\mathcal{P}(Fm)$ is a power set of $Fm$.

A consequence relation inherits all its properties from properties of
consequence operation $(1-5)$.

In our algebraic approach we identify the general notion of logic with the
structural consequence operation. The logic or deductive system in the
language of type $\mathcal{L}$ is a pair $\mathcal{S=}\left\langle \mathbf{Fm%
},\vdash _{\mathcal{S}}\right\rangle $ where $\mathbf{Fm}$ is the algebra of
formulae of type $\mathcal{L}$ and $\vdash _{\mathcal{S}}$ is a
substitution-invariant consequence relation on $\mathbf{Fm}$, that is,
relation $\vdash _{\mathcal{S}}\subseteq \mathcal{P}(Fm)\times Fm$
satisfying the conditions $(1-5).$ A logic $\mathcal{S}$ is said to be
finitary when its consequence relation satisfies the relational form of
property $(5)$, that is, when for every $\Gamma \cup \{\varphi \}\subseteq
Fm $:%
\begin{equation*}
\text{If }\Gamma \vdash _{\mathcal{S}}\varphi \text{ then there is a finite }%
\Gamma ^{\prime }\subseteq \Gamma \text{ such that }\Gamma ^{\prime }\vdash
_{\mathcal{S}}\varphi .
\end{equation*}

In our article we restrict ourselves only to finitary logics.

An identification of the notion of logic with the notion of structural
consequence operation points out in one-to-one correspondence the set of all
theories, which can be defined on the set of formulae. The sets of the form $%
X=Cn(X)$ are called theories or deductive systems. On a fixed set of
formulae - $Fm$ - one can define many different structural consequence
operations. The set of all structural consequence operations form a complete
lattice.

Based on Dishkant's work, we treat the language of quantum logics as a free
algebra \cite{Dis}. In the literature dealing with quantum logics, there
exist two different notions of logical consequence. They are determined by a
class of orthomodular lattices. The first introduced notion of logical
consequence in quantum logic is created by Kalmbach \cite{Kal}. A sentence $%
\alpha $ is a \textit{weak }logical consequence of the set $X$ of sentences
if and only if in every model and every valuation in which, every sentence
of the set $X$ has a unit of certain orthomodular lattice as its logical
value, the sentence $\alpha $ has the unit as its logical value, too.

In 1974, Goldblatt introduced the notion of \textit{strong} quantum logical
consequence: sentence $\alpha $ is a strong logical consequence of the set
of sentences $X$ if and only if for any orthomodular lattice $OML$ and any
valuation $v$, $v(\beta )\leq v(\alpha )$ for every $\beta $ $\in X$ $($the
symbol $\leq $ denotes the lattice order of $OML)$ \cite{Gol}.

All above concepts of quantum logical consequence presuppose the notion of
the model of quantum logics.

\bigskip

$\mathbf{4}$. \textbf{MODELS OF QUANTUM LOGICS.}

\bigskip

In our investigation, we employ the general method of constructing the
models of sentential calculus. We use the so-called \textit{matrix method},
which allows us to give a full algebraic description of quantum logics \cite%
{Woj,Woj1}.

By a logical matrix we mean a couple $\mathcal{M=}\left\langle \mathbf{A}%
,F\right\rangle $ where $\mathbf{A}$ is an algebra of the same similarity
type as the algebra of terms of considered sentential language and $F$ is a
subset of $A$ called the set of designated elements of $\mathcal{M}$. As
indicated we rule out neither that the set of \textit{designated elements} $%
F=\varnothing $ nor that $F=A$. The matrices of the form $\mathcal{M=}%
\left\langle \mathbf{A},\varnothing \right\rangle $ and $\mathcal{M=}%
\left\langle \mathbf{A},A\right\rangle $ are referred to as trivial. The
general concept underlying the notion of logical matrix is that the algebra
of matrix $\mathbf{A}$ is similar to the algebra of formulae of a given
propositional language. In our case, the algebra $\mathbf{A}$ is similar to
the algebra of terms of quantum logics in the sense of Dishkant \cite{Dis}.
Such logical matrix can be understood as an algebraic \textit{semantical }%
model of the considered language or simply as algebraic \textit{semantics}
for quantum logics.

The set $A$ can be considered as a \textit{range of variability} of
propositional variables. This set can be regarded as a set of \textit{%
semantical correlates} of sentential variables $($or alternatively as a set
of algebraic correlates of sentential variables$)$ \cite{Woj,Woj1}. The
concept of logical matrices regarded as models for sentential logics is of
particular importance. Every logical matrix consists of an algebra, which is 
\textit{homomorphic} with the algebra of terms of a given sentential
language. Logical matrices associated with quantum logics are formed of a
variety of $\mathbf{OL}$ or $\mathbf{OML}$. These are \textquotedblleft
natural\textquotedblright\ classes of homomorphic algebras forming logical
matrices. There are many open questions as to whether other algebras, e.g., $%
C^{\ast }$\textit{-algebras}, \textit{von Neumann algebras}, \textit{Jordan
algebras} or \textit{Grassmann algebra}, can form logical matrices for the
algebra of terms of quantum propositions. The above hints can be understood
as a link between purely logical considerations concerning bases of quantum
theory and mathematical investigations aiming at finding an appropriate
algebraic structures describing quantum reality. In this paper we restrict
ourselves only to the most natural\ algebraic semantics for quantum logics,
i.e., the variety of $\mathbf{OL}$ and $\mathbf{OML}$.

Each formula $\varphi $ of the language of quantum logic has a unique 
\textit{interpretation} in $\mathbf{A}$ depending on the value in $\mathbf{A}
$ that are assigned to its variables.

Based on the facts that $\mathbf{Fm}$ is absolutely freely generated by a
set of variables $($the set of free generators$)$ and that $\mathbf{A}$ is
an algebra of the same similarity type as $\mathbf{Fm}$, there exist a
function $f:Var\rightarrow A$ and exactly one function $h^{f}:Fm\rightarrow
A $, which is the extension of the function $f$, i.e., $h^{f}(p)=f(p)$ for
each $p\in Var$. This function is the homomorphism from the algebra of
formulae into the algebra $A$ of the logical matrix $\mathcal{M}$ = $%
\left\langle \mathbf{A},F\right\rangle $ . The set of all such homomorphisms
is denoted by $Hom_{\mathcal{S}}(\mathbf{Fm},\mathbf{A})$.

Now we can identify the interpretation of a given formula $\varphi $ of $Fm$
with $h(\varphi )$ where $h$ is a homomorphism from $\mathbf{Fm}$ to $%
\mathbf{A}$ that maps each variable of $\varphi $ into its assigned value. A
homomorphism whose domain is the algebra of terms is called an \textit{%
assignment}. One can alternatively write a formula $\varphi $ in the form $%
\varphi (x_{0},...,x_{n-1})$ to indicate that each of its variables occurs
in the list $x_{0},...,x_{n-1}$ and we write $\varphi ^{\mathbf{A}%
}(a_{0},...,a_{n-1})$ for $h(\varphi )$ where $h$ is any assignment such
that $h(x_{1})=a_{i}$ for all $i<\omega $. Given a quantum logic $\mathcal{S}
$ in a language of type $\mathcal{L}$, an $\mathcal{L}$-matrix $\mathcal{M=}%
\left\langle \mathbf{A},F\right\rangle $ is said to be a model of $\mathcal{S%
}$ if for every $h\in Hom_{\mathcal{S}}(\mathbf{Fm},\mathbf{A})$ and every $%
\Gamma \cup \{\varphi \}\subseteq Fm$%
\begin{equation*}
\text{if }h[\Gamma ]\subseteq F\text{ and }\Gamma \vdash _{\mathcal{S}%
}\varphi \text{ then }h(\varphi )\in F.
\end{equation*}

In this case it is also said that $F$ is a deductive filter of $\mathcal{S}$
or, as is common now, an $\mathcal{S}$-filter of $\mathbf{A}$ \cite{Woj1,Fo}%
. Given an algebra $\mathbf{A}$ of similarity type $\mathcal{L}$, the set of
all $\mathcal{S}$-filters of $\mathbf{A}$, which is denoted by $Fi_{\mathcal{%
S}}\mathbf{A}$ is closed under intersection of an arbitrary family and is
thus a complete lattice \cite{Fo}. By $h\in Hom_{\mathcal{S}}(\mathbf{Fm},%
\mathbf{A})$ we mean an homomorphism from the algebra of terms into the
algebra forming the logical matrices for quantum logics. Given any set of
formulae $X\subseteq A$, there is always the least $\mathcal{S}$-filter of $%
\mathbf{A}$ that contains $X$. It is called the $\mathcal{S}$-filter of $%
\mathbf{A}$ generated by $X$ and is denoted by $Fi_{\mathcal{S}}^{\mathbf{A}%
}(X)$. The class of all matrix models of quantum logic $\mathcal{S}$ is
denoted by $\mathbf{Mod}\mathcal{S}$ or $\mathbf{K}$.

Every logical matrix points out to a set of its own tautologies i.e., a set
of formulae such that $h(\alpha )\in F$ for $\alpha $ $\in Fm$ for every
homomorphisms $h\in Hom_{\mathcal{S}}(\mathbf{Fm},\mathbf{A})$. The set of
all tautologies of given matrices is denoted by $E(\mathcal{M})$. It is
invariant with respect to the endomorphisms of the algebra of terms. Every
invariant set of formulae $X\subseteq Fm$ may be represented as $E(\mathcal{M%
})=X$ with an appropriate matrix $\mathcal{M}$. The above is the well known
as \textit{Lindenbaum's theorem} \cite{Los}. For the purpose of its proof it
is enough to consider the matrix of the form $\mathcal{M=}\left\langle 
\mathbf{Fm},X\right\rangle $ . The matrices of this form are termed \textit{%
Lindebaum's matrices}. For such matrices the valuations are simply
endomorphisms of $Fm$ \cite{Los} .

Also every logical matrix determines a so-called \textit{matrix consequence
operation - }$C_{\mathcal{M}}$. For arbitrary $X\subseteq Fm$%
\begin{equation*}
C_{\mathcal{M}}(X)=\dbigcap \left\{ h^{-1}(F):h(X)\subseteq F,h\in Hom_{%
\mathcal{S}}(\mathbf{Fm},\mathbf{A})\right\}
\end{equation*}

or equivalently: for arbitrary $X\subseteq Fm$ and for arbitrary formula $%
\alpha \in Fm:\alpha \in C_{\mathcal{M}}(X)\leftrightarrow $ for every $h\in
Hom_{\mathcal{S}}(\mathbf{Fm},\mathbf{A})$ if $h(\beta )\in F$ for every $%
\beta \in X$ then $h(\alpha )\in F$.

For every matrix the operation defined in such a manner is a \textit{%
structural} and \textit{uniform} consequence. We call it the
matrix-consequence $(C_{\mathcal{M}}$, \cite{Los}$)$.

In opinion of many logicians, the above statements present the nearest
connection between sentential logics and interpretations by matrices \cite%
{Los}.

We ask what is the relationship between structural consequence operation
defined in Section 3, particularly strong and weak quantum logical
consequence and the so-called matrix consequence. We present here the
theorem $($without proof, see \cite{Los,Woj1}$)$ establishing the
conditions, which must be satisfied in order to $Cn=C_{\mathcal{M}}$.

\bigskip

\textbf{Theorem }$\mathbf{3}$ \cite{Los,Woj1}.\textit{\ Let }$Cn$\textit{\
be structural consequence operation (logic). Then }$Cn$\textit{\ is a matrix
consequence if and only if }$Cn$\textit{\ is absolutely uniform.}

\bigskip

We call a consequence $Cn$ \textit{uniform} if and only if for all set of
formulae $X,Y\subseteq Fm$ and for a formula $\alpha $ $\in Fm$, the
following conditions are satisfied:\bigskip

$1)$ $Var(X,\alpha )\cap Var(Y)=\varnothing ,$

$2)$ $Var(Y)\neq Fm,Fm$ being the set of all formulae,

$3)$ $\alpha \in Cn(X\cup Y)\bigskip $

then\bigskip

$4)$ $\alpha \in Cn(X).\bigskip $

The symbol $Var(X)$ means all free sentential variables of the set of
formulae $X$. The equation $Var(X,\alpha )\cap Var(Y)=\varnothing $ means
that the set $(X\cup \{\alpha \})$ and $Y$ have no variables in common.

The logic $Cn$ is said to be \textit{separable} if and only if given two
sets of formulae $X,Y$ of the language of $Cn$ such that $Var(X)\cap Var(Y)=$
$\varnothing $ and given any variable $r\notin Var(X\cup Y)$ the following
condition is satisfied:%
\begin{equation*}
\text{If }r\in Cn(X\cup Y)\text{ then either }r\in Cn(X)\text{ or }Cn(Y).
\end{equation*}

The separability condition can take the following stronger form.

A consequence $Cn$ will be said to be \textit{absolutely separable} if and
only if for each family $R$ of sets of formulae such that for any two sets $%
X,Y\in R$ if $X\neq Y$ then $Var(X)\cap Var(Y)$ $=$ $\varnothing $ and for
each propositional variable $r\notin Var(\dbigcup R)$%
\begin{equation*}
\text{If }r\in Cn(\dbigcup R)\text{ then }r\in Cn(X)\text{ for some }X\in R.
\end{equation*}

A consequence that is both \textit{uniform} and \textit{absolutely separable}
will be called \textit{absolutely uniform}.

The logical matrices determining consequence operation, which is equal to
the structural consequence operation, i.e., $Cn=C_{\mathcal{M}}$, are called 
\textit{strongly adequate} logical matrices \cite{Woj1}.

As it is stated in Section $3$ in the language of quantum logic, we can
define two different consequence operations: the weak one and the strong one.

The strong consequence operation is determined by the class of models of
quantum logic as follows:%
\begin{equation*}
\Gamma \vdash _{\mathcal{S}}\varphi \text{ iff }\forall \mathbf{A\in
OML,\forall }h\in Hom_{\mathcal{S}}(\mathbf{Fm},\mathbf{A})\text{ }\forall
a\in A\text{ }
\end{equation*}%
\begin{equation*}
\text{if }a\leq h(\beta )\text{ }\forall \beta \in \Gamma \text{ then }a\leq
h(\varphi ).
\end{equation*}

The weak consequence operation is determined by the class of models of
quantum logic as follows:%
\begin{equation*}
\Gamma \vdash _{\mathcal{S}}\varphi \text{ iff }\forall \mathbf{A\in OML}%
,\forall h\in Hom_{\mathcal{S}}(\mathbf{Fm},\mathbf{A})\text{ if }h(\beta )=%
\mathbf{1}\text{ }\forall \beta \in \Gamma \text{ then }h(\varphi )=\mathbf{1%
}.
\end{equation*}

The names \textquotedblleft weak\textquotedblright\ and \textquotedblleft
strong\textquotedblright\ are misleading because the weak quantum
consequence operation is the strengthening of the strong quantum consequence
operation \cite{Mal}. These names persist only from historical reasons.

In the above formal exposition of the two different definitions of quantum
logical consequence, we consider an algebra $\mathbf{A}$ as belonging to the
variety of $\mathbf{OML}$. Based on the definition of quantum logical
consequence we can uniquely point out the classes of algebras constituting
the matrix (algebraic) semantics for quantum logics.

\bigskip

\textbf{Corollary} $\mathbf{4}$. \textit{The class of matrices}%
\begin{equation*}
\mathbf{Mod}\mathcal{S}=\left\{ (\mathbf{A},[a)):\mathbf{A}\in \mathbf{Mod}%
\mathcal{S},a\in A\right\}
\end{equation*}

\textit{is a matrix semantics for the strong version of quantum logic. }$[a)$%
\textit{\ is a principal filter of the form }$\left\{ x\in A:x\geq a\right\} 
$\textit{.}

\bigskip

\textbf{Corollary} $\mathbf{5}$. \textit{The class of matrices}%
\begin{equation*}
\mathbf{Mod}\mathcal{S=}\left\{ (\mathbf{A},\{1\}):\mathbf{A}\in \mathbf{Mod}%
\mathcal{S}\right\}
\end{equation*}

\textit{is a matrix semantics for the weak version of quantum logic where
the filter is of the form }$F=\{1\}.$

\bigskip

$\mathbf{5}$. \textbf{CONCLUSION.}

\bigskip

In our paper, we did not consider any physical implications of different
forms of quantum logical consequence operations. Following the main idea
that any logic can be understood as a structural consequence operation, we
indicated adequate semantics for quantum logics. Investigations carried out
in this paper consist first report concerning more general topic -
\textquotedblleft \textit{Inference in Quantum Logics}\textquotedblright .
We plan to present consequence operation define on \textit{Greechie diagram}%
. In order to do that, we will introduce the notion of \textit{Greechie
diagram satisfiability}. Results of these investigations will be presented
elsewhere.

There are also reports treating consequence operation in quantum logics as a
kind of \textit{nonmonotonic reasoning} \cite{Eng}. Above approach will be
confronted with our statements considering consequence operation in quantum
logics as a kind of monotonic reasoning.

\bigskip

\bigskip\ 

\end{document}